\documentclass[%
 reprint,
superscriptaddress,
showpacs,
preprintnumbers,
nofootinbib,
nobibnotes,
 amsmath,amssymb,
 aps,
]{revtex4-1}

\usepackage{graphicx}% Include figure files
\usepackage{dcolumn}% Align table columns on decimal point
\usepackage{bm}% bold math

\usepackage{amsfonts}
\usepackage{upgreek}     % for non italic greek symbols
\usepackage[T1]{fontenc} % for accents etc
\usepackage{ae,aecompl}  % for pdf fonts

\newcommand{\cm}{\ensuremath{~\mathrm{cm}^{-1}}}
\newcommand{\micron}{~\ensuremath{\upmu\text{m}}}
\newcommand{\microsec}{~\ensuremath{\upmu\text{s}}}

\begin{document}

\title{Driving rotational transitions in molecules on a chip}
\author{Gabriele Santambrogio}
\email{gabriele.santambrogio@fhi-berlin.mpg.de}
\author{Samuel A. Meek}
\author{Mark J. Abel}
\affiliation{Fritz-Haber-Institut der Max-Planck-Gesellschaft,
Faradayweg 4-6, 14195 Berlin, Germany}
\author{Liam M. Duffy}
\affiliation{Dept.~of Chemistry and Biochemistry,
The University of North Carolina at Greensboro, 310 McIver St., Greensboro NC, USA}
\author{Gerard Meijer}
\affiliation{Fritz-Haber-Institut der Max-Planck-Gesellschaft,
Faradayweg 4-6, 14195 Berlin, Germany}
\date{\today}
\pacs{37.10.Pq, 33.20.Bx, 33.57.+c, 37.20.+j, 
42.15.Dp, 42.50.Md}

\maketitle

\section*{Abstract}
Polar molecules in selected quantum states can be guided, decelerated and trapped 
using electric fields created by microstructured electrodes on a chip. Here we explore 
how transitions between two of these quantum states can be induced while the
molecules are on the chip. We use CO ($a^3\Pi_1, v=0$) molecules, prepared in
the $J=1$ rotational level, and induce the $J=2$ $\leftarrow$ $J=1$ rotational 
transition with narrow-band sub-THz (mm-wave) radiation. First, the mm-wave source is 
characterized using CO molecules in a freely propagating molecular beam, and
both Rabi cycling and rapid adiabatic passage are examined. Then, we demonstrate 
that the mm-wave radiation can be coupled to CO molecules that are less than 
50\micron\ above the chip. Finally, CO molecules are guided in the $J=1$ level 
to the center of the chip where they are pumped to the $J=2$ level, recaptured, and 
guided off the chip.

\section{Introduction}
The manipulation of polar molecules above a chip using electric fields 
produced by microstructured electrodes on the chip surface is a 
fascinating new research field \cite{Meek:2009p1699,Meek:2009p012063}. 
Miniaturization of electric field structures
enables the creation of large field gradients, i.e. large forces and tight 
potential wells for polar molecules. Above a chip, the positions of these 
potential wells, and thereby the positions of the trapped molecules, can 
be controlled to an extreme precision. In addition, present-day 
microelectronics technology makes it possible to integrate multiple tools 
and devices onto a compact surface area. These can include lenses, 
decelerators and traps for polar molecules but also integrated detection 
elements like radiation sources and optical cavities, for instance. Over a 
decade ago, similar notions launched the field of atom
chips\cite{Weinstein:1995p4004,Hinds:1999pR119,Fortagh:2007p235},  
in which atoms are manipulated and controlled above a chip using magnetic fields 
produced by current-carrying wires. Whereas atom chips have been used to
demonstrate rapid Bose-Einstein condensation\cite{Haensel:2001p498} and have
already found applications in matter-wave interferometry\cite{Schumm:2005p57} and
in inertial and gravitational field sensing\cite{Zoest:2010p1540}, molecule chips
are still in their infancy and their potential still needs to be explored.  

A particular advantage of using molecules instead of atoms on a chip is 
that they can be coupled to photons over a wider range of frequencies.
The fundamental molecular vibrational modes can be addressed with 
mid-infrared photons whereas their overtones and combination modes 
extend into the near-infrared range. In addition, polar molecules have a 
dense spectrum of rotational transitions in the sub-THz, or mm-wave, 
region of the spectrum. During the last years, several schemes have been 
proposed for quantum computation in which trapped polar molecules in 
selected quantum states serve as qubits.\cite{Andre:2006p636,Demille:2002p067901}
In these schemes, flipping of a 
qubit can be accomplished by inducing a transition to another internal 
quantum state in the molecule, provided the molecule remains trapped in
the final state as well. Transitions between rotational levels within a given 
electronic and vibrational state are especially well suited for this; pairs
of rotational levels that can be trapped in external fields and that are 
connected via electric dipole allowed transitions can be found. Transitions 
between these rotational levels can be driven with unit efficiency and 
yet the lifetimes of the rotationally excited levels are very long; vibrational 
and electronic transitions, in contrast, often result in leakage of population 
out of the system of coupled states. Moreover, the required narrow-band, 
coherent radiation sources spanning the microwave to sub-THz frequency 
range can be integrated on a microchip and no adverse effect of the 
microwave radiation on the performance of the chip components is expected. 

In the experiments reported here we explore the use of 
rotational transitions in molecules on a chip. The system we use is the carbon 
monoxide molecule, prepared with a pulsed laser in a single rotational 
level ($J=1$) of its first electronically excited, metastable state ($a^3\Pi_1, v=0$). 
The $J=2$ $\leftarrow$ $J=1$ rotational transition is induced using a continuous 
wave source of radiation with a wavelength of around 1.5 mm. Molecules
in the rotationally excited $J=2$ level are subsequently state-selectively 
detected using ionization with another pulsed laser system. 

The paper consists of two parts. In the first part, the experimental
set-up is described and the mm-wave source as well as the rotational 
excitation step are characterized using experiments on CO molecules 
in a freely propagating molecular beam. These experiments demonstrate 
that efficient and robust rotational excitation can be achieved when the 
molecules pass through the focused mm-wave beam at the appropriate 
position, such as to fulfill the conditions for rapid adiabatic passage. In 
the second part, experiments are described in which the mm-wave 
radiation interacts with the metastable CO molecules while these are 
propagating with a constant velocity at a close distance above the chip. 
These experiments explicitly demonstrate that the mm-wave radiation
can be coupled onto the chip and that the CO molecules --- guided in 
moving electric field traps to the center of the chip while in the $J=1$ 
level --- can be pumped to the $J=2$ level while on the chip. The 
rotationally excited molecules can subsequently be re-captured in the 
miniaturized electric field traps, transported to the end of the chip and 
detected. 

\section{Rotational excitation of metastable CO in a beam}
\begin{figure}
\centering
\includegraphics[width=0.47\textwidth]{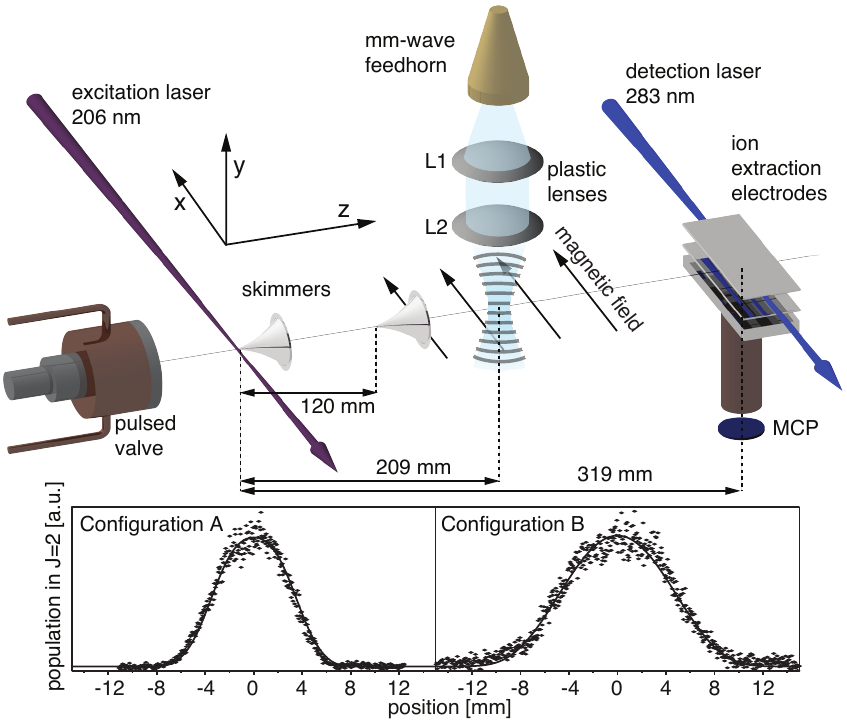}
\caption{Scheme of the molecular beam setup, with the distances along
the molecular beam line indicated (not drawn to scale). CO molecules are
prepared in selected rotational levels of the $a^3\Pi_1, v=0$ state
using 206 nm radiation, interact with mm-wave radiation that is resonant 
with rotational transitions in the metastable state and are subsequently
state-selectively detected via resonant ionization. In the lower part, the 
measured (dots) and fitted (solid curve) spatial profile of the mm-wave 
beam is shown around the waist (Configuration A) and in the divergent 
part of the beam (Configuration B).
\label{F:BLFree}} 
\end{figure}

To characterize the mm-wave source and the rotational excitation 
scheme in metastable CO, we first performed experiments using 
the setup that is schematically depicted in Figure~\ref{F:BLFree}. A mixture of 
20\% CO in Krypton is expanded into vacuum from a pulsed valve 
(General Valve; series 99) operating at a repetition rate of 10~Hz. 
Before passing through the first skimmer, the CO molecules are excited 
from the $N''=1$ rotational level in the electronic and vibrational 
ground-state ($X^1\Sigma^+, v"=0$) to the upper
$\Lambda$-doublet component of the $a^3\Pi_1, v=0, J=1$ level. This
particular level has a radiative lifetime of 2.6~ms.\cite{Gilijamse:2007p1477}
Alternatively, the ground state molecules can be excited from the $N''=2$
level to the upper $\Lambda$-doublet component of the $J=2$ level. 
The 206~nm radiation from the pulsed excitation laser (1~mJ in a 
5~ns pulse with a bandwidth of about 150~MHz) is weakly focused 
onto the molecular beam.
In this way, an approximately 1~mm$^3$ packet containing about 
10$^8$ metastable CO molecules is obtained after the skimmer.
With the pulsed valve at room 
temperature, the mean velocity of the CO molecules is around 470~m/s 
and the velocity spread is rather large, with a full width at half maximum of
80~m/s.

To better collimate the molecular beam, the CO molecules pass through a
second 1~mm diameter skimmer on their way to the 
interaction region with the mm-wave radiation. The mm-waves are produced outside
of vacuum. A microwave synthesizer 
(\emph{Agilent}, model HP83623B with option 008; 1~Hz resolution)
drives an ``Armadillo'' millimeter-wave source module (\emph{Agilent},
model 83558A, 75--110~GHz, $\sim$0~dBm output)  which in turn pumps
a millimeter-wave power amplifier (\emph{Spacek Labs Inc.}, model
SPW-18-14; 75--110~GHz, $\sim$+10~dB).  The output of this W-band
amplifier is then frequency doubled with a passive multiplier
(\emph{Virginia Diodes Inc.}, models VDI-WR5.1x2; 150--220~GHz), losing
roughly 12~dB and resulting in a mm-wave beam of roughly $-$2~dBm or 0.6~mW
at 200~GHz.  A standard gain horn from \emph{Millitech} (model
SGH-05-RC000) is used to launch the mm-wave beam into free
space.\cite{Duffy:2005p093104} While the use of a corrugated scalar
feedhorn would have ensured a pure Gaussian beam profile, it is evident from the measured
profiles in Figure~\ref{F:BLFree} that the presence of side-lobes is within the
noise level of the experiment. The mm-wave beam waist at the exit of the horn is
then expanded and transfered to the molecular beam with a pair of plastic lenses
(L1 Teflon, L2 TPX) configured in a Gaussian beam telescope
geometry.\cite{Goldsmith:1982p227} The lenses' focal lengths are 6 and 15~cm,
respectively. The mm-wave beam
is introduced into the vacuum chamber through a Teflon window and intersects the
molecular beam under right angles. In this experiment, the Armadillo is oriented
so that the polarization of the mm-waves is parallel to the molecular 
beam axis and is set to drive 
the $J=2 \leftarrow J=1$ transition in the $a^3\Pi_1, v=0$ state
CO. Since the Earth's magnetic field induces Zeeman splittings 
of the rotational levels, an offset magnetic field of about 200~Gauss is
applied to define the main direction of the magnetic field to be perpendicular 
to the polarization of the mm-radiation. In
this configuration, the selection rules for the various $M$-components of
the rotational $J=2 \leftarrow J=1$ transition are $\Delta M=\pm 1$.

\begin{figure}
\centering
\includegraphics[width=0.47\textwidth]{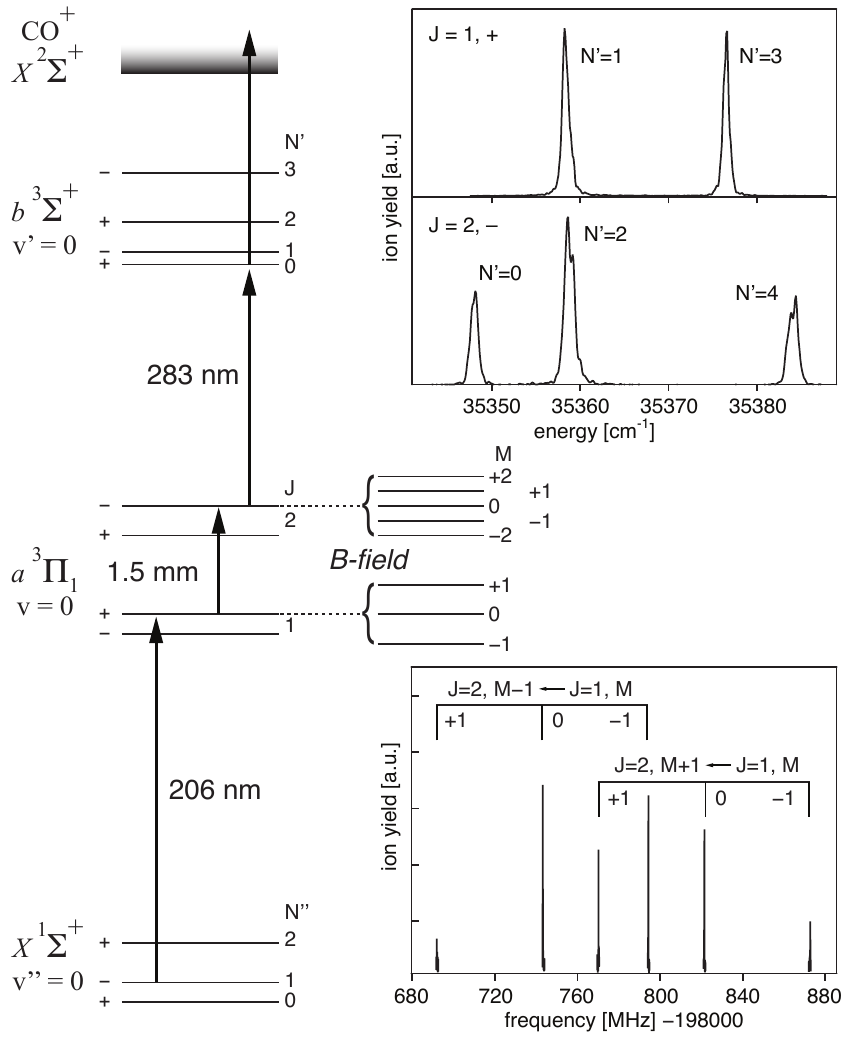}
\caption{Scheme of the energy levels of CO involved in the preparation
of the molecules in the metastable state, in the rotational transitions in
this state, and in the subsequent resonance enhanced ionization detection.
The (1+1)-REMPI spectrum starting from two different $J$ levels in the
metastable state is shown, together with the $M$-resolved 
$J=2 \leftarrow J=1$ transition in a magnetic field of 200 Gauss.
\label{F:CO-levels}} 
\end{figure}

Downstream from the mm-wave interaction zone, the 
metastable CO molecules can be state-selectively detected. Carbon
monoxide molecules in the $a^3\Pi_1, v=0$ state can be resonantly
excited to selected rotational levels of the $b^3\Sigma^+, v'=0$ state,
from which they can be ionized by another photon from the 
same laser pulse. A tunable pulsed laser (4~mJ energy in a 5~ns pulse with 
a bandwidth of 0.2\cm) operating around 283~nm is used for
this (1+1) resonance enhanced multi-photon ionization (REMPI) detection
scheme. The ionization laser beam intersects the molecular beam 
perpendicularly and is only weakly focused with a diameter of about 
1~mm. The resulting parent ions are mass-selectively detected in a 
compact linear time-of-flight (TOF) mass-spectrometer equipped with
a micro-channel plate (MCP) detector. Ionization is performed with the
voltages on the ion extraction electrodes switched off in order to
avoid parity-mixing of the $\Lambda$-doublet components in the 
metastable state. A scheme of the energy levels involved in the preparation
of the metastable molecules, in the mm-wave excitation and in the 
subsequent (1+1)-REMPI detection is shown in Figure~\ref{F:CO-levels}. 
In this Figure, the measured (1+1)-REMPI spectra from the upper 
$\Lambda$-doublet components of the $J=1$ and $J=2$ levels are 
shown as well.  

If the metastable CO molecules are prepared in the $J=1$ level while 
the $J=2$ level is probed by (1+1)-REMPI, a mm-wave 
spectrum of the $J=2 \leftarrow J=1$ transition can be recorded against 
zero background. Such a spectrum, measured with approximately 20~$\upmu$W/cm$^2$
of mm-wave radiation, is shown in Figure 2 as well. 
Three $\Delta M = -1$ and three $\Delta M = +1$ transitions are observed. 
The observed width of the individual transitions of about 50~kHz is 
attributed to the limited interaction time of the CO molecules with the 
mm-wave radiation (a few tens of microseconds) in combination with 
the slight inhomogeneity of the applied magnetic field. 

\subsection{Characterizing the mm-wave beam}

The combination of the laser preparation of metastable CO 
at a well-defined time and position with state-selective 
detection at a known delay and distance further downstream 
enables us to selectively monitor molecules within a narrow velocity interval.
The doubly-skimmed molecular beam has a full width at half maximum 
transverse spread of approximately 4~mm in the ionization detection region. 
With the experimental geometry as shown in Figure~\ref{F:BLFree}, only 
metastable CO molecules in a 1~mm diameter and 4~mm 
long cylinder, oriented with its long axis perpendicular to the molecular 
beam and perpendicular to the propagation direction of the mm-wave 
radiation, are detected. In the mm-wave excitation zone, the length
of the cylindrical volume of the metastable CO molecules that will be
probed further downstream is only 2.5~mm. This experimental setup enables 
an accurate {\it in situ} characterization of the spatial profile of the mm-wave  
radiation. For this, metastable CO molecules are prepared in the $J=1$ 
level and probed from the $J=2$ level. The mm-wave radiation, polarized
parallel to the external magnetic field in this case, is kept 
fixed to the center frequency of the $J=2, ~M=0 \leftarrow J=1, ~M=0$ 
transition at 198782.250~MHz. The preparation and detection laser are timed such that only 
molecules with a velocity of 470$\pm$1~m/s are detected. The mm-wave 
radiation is only switched on during a 2\microsec\ time interval. By measuring
the (1+1)-REMPI signal from the $J=2$ level as a function of the time delay
between laser preparation and the center of the 2\microsec\ interval, the
spatial profile of the mm-wave radiation is sampled. The result of such scans
is shown in the lower part of Figure~\ref{F:BLFree}, for two different positions 
of the beam waist; the measured (dots) spatial profile is shown when the waist of
the mm-wave beam is close to the axis of the molecular beam (Configuration A) and
when the waist is moved closer to the mm-wave source, i.e., when the molecules 
cross a divergent mm-beam (Configuration B).

For a Gaussian beam with peak amplitude $\mathcal{E}_0$ and waist size $w_0$,
the complex electric field amplitude is given by 
\begin{multline}\label{eq:gauss}
\mathcal{E}(\rho,y)=\mathcal{E}_0\frac{w_0}{w(y)}\exp\Bigl[\frac{-\rho^2}{w^2(y)}\Bigr]\\
\exp\Bigl[-i\frac{2\pi y}{\lambda}-i\frac{\pi\rho^2}{\lambda r(y)}+
i\arctan\bigl(\frac{\lambda\,y}{\pi\,w_0^2}\bigr)\Bigr]
\end{multline}
where $\rho$ is the radial distance from the axis of the beam, $y$ is the axial
distance from the waist, $w(y)=w_0\sqrt{1+(y\lambda/(\pi\,w_0^2))^2}$ is the radius 
at which the electric field amplitude drops to $1/e$ of its value on axis, 
$r(y)=y[1+(\pi\, w_0^2/(\lambda \, y))^2]$ is the radius of curvature of the 
beam's wavefronts and $\lambda$ is the wavelength. The solid curves in panel
A and B of Figure~\ref{F:BLFree}, are fits to the calculated excitation probability
to the $J=2$ level ({\it vide infra}) using this expression for the electric field
distribution. From this fit, we extract a value of $w_0$ = 5~mm for the beam waist radius.

\subsection{Rabi oscillations and rapid adiabatic passage}
 
\begin{figure}
\centering
\includegraphics[width=0.47\textwidth]{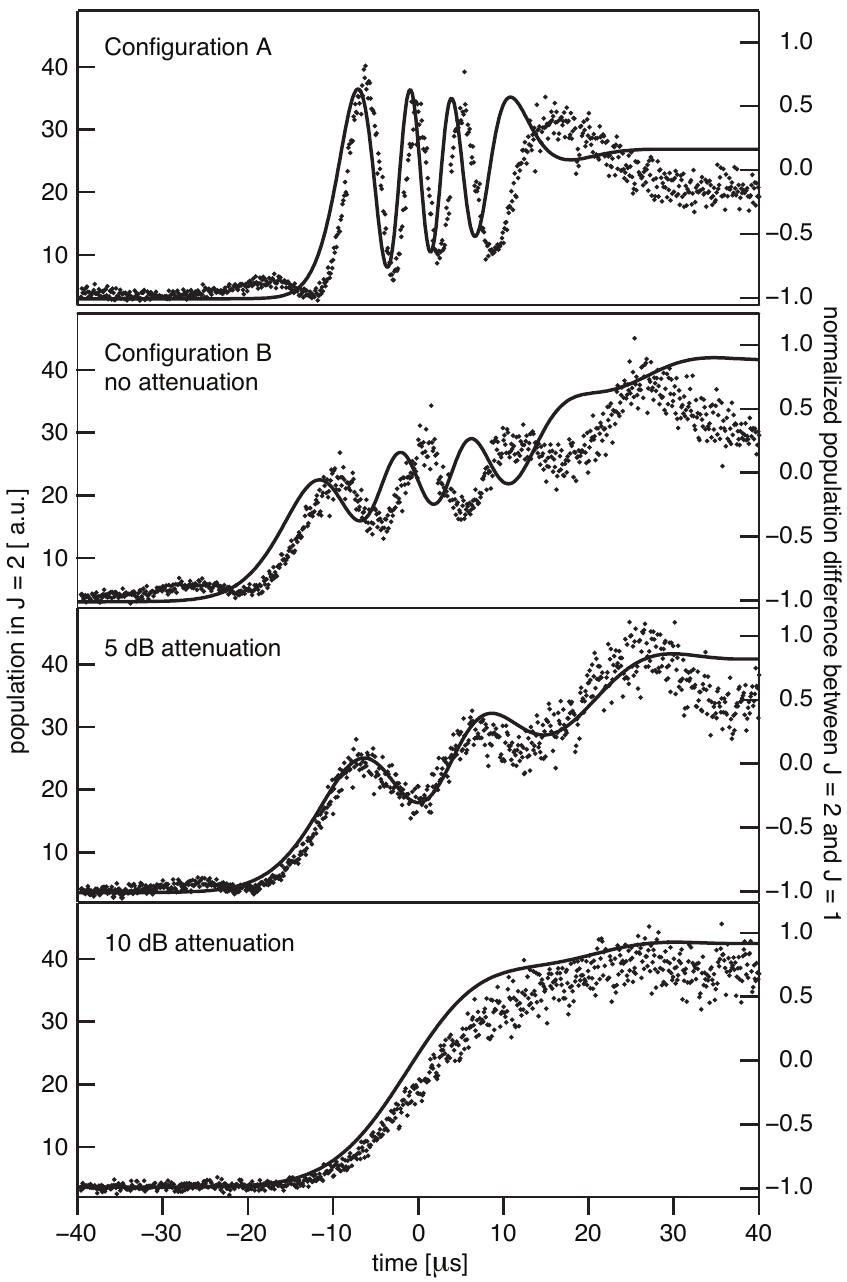}
\caption{Measured (dots) and simulated (solid curve) population in the rotationally 
excited $J=2$ level as a function of time; time zero is defined 
as the moment when the molecules are in the center of the mm-wave beam. In the upper panel, the 
molecules pass close to the waist of the mm-wave beam. In the lower three panels, 
the molecules pass through the divergent part of the mm-wave beam and thereby 
experience a frequency chirp. The ratio of the frequency chirp to the Rabi frequency 
is increased in going from the second panel via the third panel to the lowest one by 
attenuating the mm-waves. Almost complete transfer of the rotational population to 
the final $J=2$ level is obtained in this rapid adiabatic passage scheme, as indicated 
by the value of $r_3$ on the axis on the right.\label{F:rabi}}
\end{figure}

If the mm-wave radiation is tuned to the frequency of an allowed transition,
the metastable CO molecules undergo Rabi oscillations between the two 
coupled levels while they fly through the mm-beam.\cite{Rabi:1937p652,Ramsey,Cohen-Tannoudji}
Abruptly turning off the mm-wave source while the molecules are still in the 
interaction region interrupts the Rabi cycling. The CO molecules will fly on and 
when they arrive in the ionization detection region they still have the population 
distribution of the moment when the mm-wave is switched off. This enables 
us to directly monitor the population distribution as a function of time, thereby 
visualizing the Rabi oscillations. The top panel of Figure~\ref{F:rabi} shows the measured 
population in the $J=2$ level as a function of the switch-off time of the mm-wave 
source, which is tuned to the center frequency of the $J=2, ~M=0 \leftarrow J=1, ~M=0$
transition with an intensity of about 0.2~mW/cm$^2$. In these measurements
the molecules pass close to the waist of the mm-wave beam (Configuration A in
Figure~\ref{F:BLFree}); time zero is defined as the moment when the molecules
are in the center of the mm-wave beam.

To describe the evolution of the two level system, $J=1$ and $J=2$, in the presence 
of a light field $\mathcal{E}\cos(\omega t)$, we use the representation suggested 
by Feynman et al..\cite{Feynman:1957p49} The wave function of a molecule is  
\begin{equation}
\psi(t)=a(t)\psi_{J=1}+b(t)\psi_{J=2}
\end{equation}
and the energies of the two eigenstates are $E_{J=1}=\hbar\omega_a$ and
$E_{J=2}=\hbar\omega_b$. We define a so-called population vector $\vec r$
$\equiv$ ($r_1$, $r_2$, $r_3$) = ($ab^*+a^*b$, $i(a^*b-ab^*)$, $bb^*-aa^*$),
a field vector $\vec \Omega$ $\equiv$ ($\Omega_1$, $\Omega_2$, $\Omega_3$) =
($-\mu\, \mathcal{E}/\hbar$, $0$, $\omega_0-\omega$), where $\mu$ is the dipole
moment of the transition, 0.61 Debye in this specific case\cite{Meek:2010phd},
and $\omega_0\equiv\omega_b-\omega_a$. Then the time evolution of the system is described 
by the equation
\begin{equation}\label{eq:motion}
\frac{d\vec r}{dt} = \vec \Omega \times \vec r.
\end{equation}
The first two components of $\vec r$ are the real and imaginary part of the coherence
between the amplitudes in the two levels, whereas $r_3$ is the population 
difference between the upper and the lower level. The first component of $\vec \Omega$
is the angular Rabi frequency at resonance and the third component is the 
detuning from resonance. Equation~\eqref{eq:motion} states that the 
population vector $\vec r$ precesses around the field vector $\vec \Omega$. 
For instance, when the radiation field is on resonance and the initial
population is in the $J=1$ level ($\vec r= (0,0,-1)$), 
equation~\eqref{eq:motion} dictates an evolution of the population vector as 
$\vec r (t) = (0,\sin(\Omega_1 t), -\cos (\Omega_1 t))$, as expected for Rabi 
oscillations. If the frequency nor the intensity of the electric field changes over time, 
the probability to be in the rotationally excited level is 
\begin{equation}\label{eq:rabi}
|b(t)|^2=\Bigl(\frac{\Omega_1}{\Omega}\Bigr)^2\sin^2\Bigl(\frac{\Omega\,t}{2}\Bigr).
\end{equation}

It is evident that by abruptly interrupting the resonant ($\Omega=\Omega_1$)
Rabi cycling at 
$t=\pi/\Omega_1$, it is possible to transfer the whole population from one state
to the other. However, this is not a very robust method, mainly because
$\Omega_1$ depends on the electric field strength, so any variation of the field
strength requires a correction of the switch-off time. Moreover, molecules 
interacting with different parts of the mm-wave beam experience different 
field strengths and therefore cycle with different frequencies. This effect 
can actually be seen in the experimental data shown in the top panel
of Figure~\ref{F:rabi}: the amplitude of the oscillations progressively 
decreases, as molecules flying through regions with different electric field 
strengths accumulate different phases. The solid curve shown in this panel
is a simulation (not a fit) computed using the formalism described here with only
experimentally determined variables for input. The waist 
$w_0$ of the mm-wave beam and the distance from the waist where the 
molecules pass through are extracted from the fit of the spatial profiles 
shown in Figure~\ref{F:BLFree}, in which the expression for $|b(t)|^2$ at 
$t$ = 2\microsec\ has been used.    

To efficiently transfer population between two levels in a more robust way,
rapid adiabatic passage can be used.\cite{Treacy:1968p421} The fundamental  
idea behind rapid adiabatic passage is to have a small angle between the field
vector $\vec \Omega$ and the population vector $\vec r$; since $\vec r$
precesses around $\vec\Omega$ the population will follow the field vector when
this one is made to evolve slowly to the desired orientation. The degree of
control over the population transfer is limited by the angular extent of the
precession cone. When the initial population is in the $J=1$ level, i.e. $\vec r
= (0,0,-1)$, the field vector $\vec \Omega$ can be made almost parallel to the
population vector $\vec r$ by choosing a detuning that is much larger than the
Rabi frequency. Then, by sweeping the radiation frequency through the resonance
until the detuning is again much larger than the Rabi frequency but with
inverted sign, all the population is transferred to the $J=2$ level. The
direction of the frequency sweep is irrelevant for the population transfer. The
condition for the process to be adiabatic is that the rate of change in the
direction of $\vec \Omega$ is smaller than the precession frequency, which is
proportional to $\Omega$.\cite{Wichman:1990p358}

In our experiments, the frequency sweep required for the rapid adiabatic passage 
results from the Doppler shift that occurs when the metastable CO molecules pass 
through the curved wavefronts of the divergent part of the Gaussian mm-wave beam. 
This Doppler shift, $\Delta\omega(t)$, is the time derivative of the phase $\phi(t)$ 
of the electric field as seen by the molecules. Since the molecules fly along the 
$z$-axis with a velocity $v$, the Doppler shift is $(\partial \phi/ \partial z)\,v$, or, 
from Equation~\eqref{eq:gauss},
\begin{equation}\label{eq:detuning}
\Delta\omega(t)=-\frac{2\pi v^2}{\lambda r(y)}t.
\end{equation}
The condition that the initial and final detuning must be much larger than the 
Rabi frequency is automatically fulfilled when the molecules pass through 
the divergent part of the Gaussian beam as they experience a Gaussian time 
dependence of the electric field strength together with a linear frequency 
chirp\footnote{Note that if the distance between the molecular beam axis and 
the mm-beam axis is $x \neq 0$, the molecules experience a field amplitude 
reduced by a constant factor $\exp[-x^2/w^2(y)]$, but that the overall Gaussian 
time dependence and the Doppler shift do not depend on $x$. In the experiments, 
we did not optimize the $x$-position, although we expect to have been close 
to the $x=0$ position based on the relative widths and amplitudes of the data
shown in Figure~\ref{F:BLFree}; in any case, the $x$-offset remained constant throughout the
experiments.}.

In the lower three panels of Figure~\ref{F:rabi}, the measured population in the $J=2$
level is shown (dots) as a function of switch-off time of the mm-wave radiation when
the molecules pass through the divergent part of the mm-wave beam, about 4~cm
away from the beam waist (Configuration B in Figure~\ref{F:BLFree}). In going
from the second panel to the third panel, the power of the mm-waves is decreased
by about a factor three while in the lowest panel the power is a factor ten
below the one in the second panel. This reduction in mm-wave power results in
an increase of the ratio of the frequency chirp to the Rabi frequency, and thus
into a rapid adiabatic passage with ever fewer oscillations. By numerically
solving Equation~\eqref{eq:motion}, using the two fitted parameters of the mm-wave
beam and the known experimental geometry as input, the simulated curves (solid)
are obtained. On the axis on the right, the third component of the population
vector, $r_3$,  is shown from which it is seen that robust and near-complete
transfer of population to  the final $J=2$ level is obtained in this rapid
adiabatic passage scheme.

\section{Experiments on the molecule chip}
\begin{figure}
\centering
\includegraphics[width=0.47\textwidth]{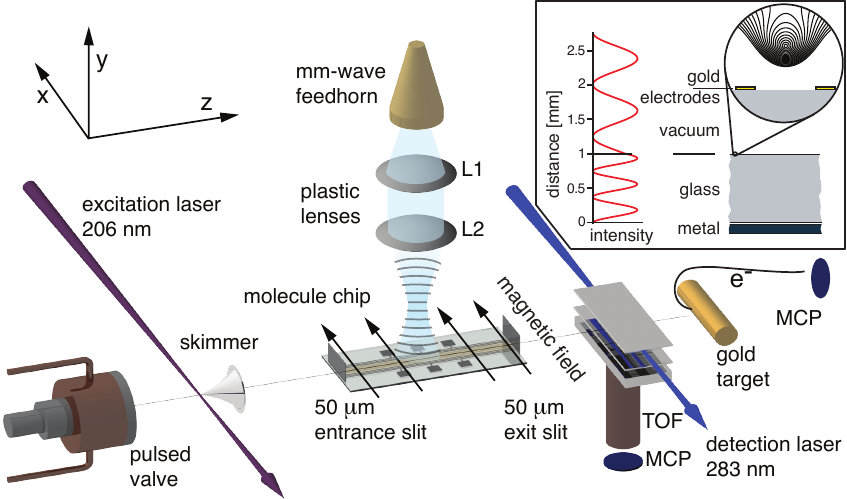}
\caption{Scheme of the experimental setup used to study mm-wave excitation of CO
molecules on a chip. Apart from detection via ionization, the metastable CO molecules 
can be detected via recording of the Auger electrons that are emitted when the molecules 
hit a gold target somewhat further downstream. In the box in the upper right part, a cross-section of the 
molecule chip is given, together with a plot of the intensity of the mm-wave radiation as 
a function of the distance from the metal holder on which the glass chip is mounted. Above 
the two microstructured gold electrodes shown in the zoom-in, equipotential lines of the 
electric field are shown, indicating an electric field minimum some 25\micron\  above the 
surface of the chip.\label{F:BLChip}}
\end{figure}

The experimental setup that is used to interrogate CO molecules on the chip with 
mm-wave radiation is shown in Figure~\ref{F:BLChip}. The chip used for these 
experiments is a new, longer version of the chip that was previously described by 
our group.\cite{Meek:2009p1699,Meek:2008p153003,Meek:2009p055024} It consists of an array of 1848 
equidistant gold electrodes, each of which is 4~mm long, 10\micron\ wide and 
approximately 100~nm high. They are deposited onto a 1~mm thick glass substrate 
with a 40\micron\ center-to-center spacing, forming a structure that is 74~mm long 
(\emph{Micro Resist Technology GmbH}). When appropriate potentials are applied to the 
electrodes, tubular electric field geometries of 4~mm length and 20\micron\ diameter 
are generated, on the axis of which the electric field strength drops to
zero. These electric field 
geometries, centered roughly 25\micron\ above the chip surface, act as 
traps for molecules in low-field-seeking quantum states. Two traps are formed for 
each 6 electrodes on the chip surface, i.e., the traps are spaced 120\micron\ 
apart. For metastable CO molecules in the low-field-seeking components of the 
$J=1$ and $J=2$ levels, these traps have a depth of 65~mK and 30~mK, respectively. 
When sinusoidal waveforms are applied to the electrodes, the traps translate
over the chip at a constant height and without changing their shape. The speed of 
the motion of the trap is directly proportional to the frequency of the waveform. 
When the frequency of the waveform is constant the traps move at constant velocity 
and the device acts as a guide. If the frequency of the waveforms is chirped down, 
the traps slow down and the chip can be used as a decelerator for low-field-seeking 
polar molecules. For more details on the guiding, deceleration and trapping of 
metastable CO molecules on the chip, the reader is referred to the existing
literature (see, in particular, ref.\cite{Meek:2009p055024}).
In the present study the molecule chip has only been used as a supersonic molecular 
conveyor belt, guiding the metastable CO molecules over the chip at a constant
velocity.

It is not \emph{a priori} obvious that the mm-wave radiation can be coupled to the molecules
on the chip, as these move only at a distance of about 1/60 of the
wavelength above the
plane of the metallic electrodes. As indicated in Figure~\ref{F:BLChip}, the mm-waves are 
coupled onto the molecule chip perpendicular to its surface. To avoid reflection of the 
radiation from the microstructured electrodes, the polarization of the mm-waves has to be 
perpendicular to the electrodes, i.e., parallel to the molecular beam axis. The mm-waves
are then largely transmitted through the electrodes, travel through the glass substrate (refractive 
index about 2)\cite{Lamb:1996p1997}, and reflect from the steel plate on which the glass chip is mounted. 
As schematically shown in the inset of Figure~\ref{F:BLChip}, the stationary electric 
field intensity of the mm-wave radiation is expected to be about 75\% of its maximum value 
at the location of the molecules.

\begin{figure}
\centering
\includegraphics[width=0.47\textwidth]{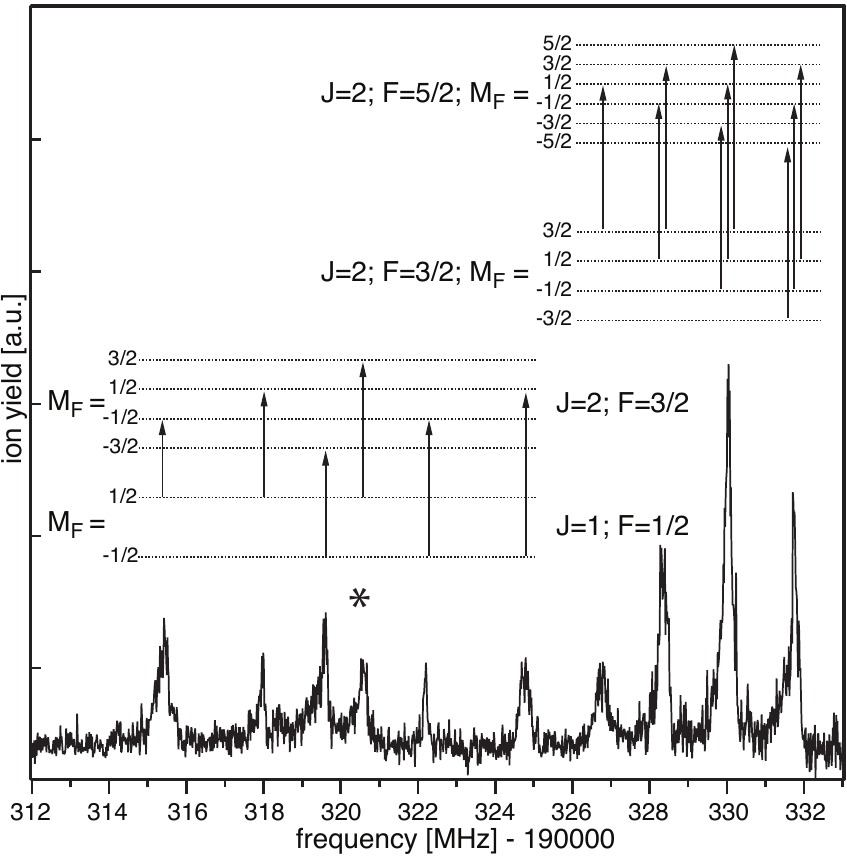}
\caption{Rotational $J=2 \leftarrow J=1$ transition of $^{13}$CO molecules in the metastable 
$a^3\Pi_1, ~v=0$ state in an external magnetic field of 10~Gauss. The spectrum is induced 
above the chip, with the polarization of the mm-wave radiation perpendicular to the direction
of the magnetic field, and is recorded by detecting the molecules in the $J=2$ level via 
(1+1)-REMPI. The observed $M$-resolved transitions are assigned in the energy level scheme 
directly above the spectrum. The peak marked with an asterisk is the transition
used for the experiment in section~\ref{sec:cleaning}.\label{F:13CO}}
\end{figure}

To explicitly demonstrate that the mm-wave radiation can be coupled onto the
chip and that rotational transitions can be induced in molecules at close
distance above the chip surface, we measured the $J=2 \leftarrow J=1$ rotational
transition in CO molecules above a chip. We used $^{13}$CO molecules for these
measurements because, as will be shown later, these can be guided more
efficiently on the chip; the nuclear spin of the $^{13}$C atom and the resulting
hyperfine splitting of the rotational levels in $^{13}$CO prevents losses due to
non-adiabatic transitions in this isotopologue.\cite{Meek:2009p1699} An external
magnetic field of 10~Gauss, perpendicular to the polarization direction of the
mm-wave radiation, is applied and the $M$-resolved rotational spectrum is
recorded by detecting the molecules in the $J=2$ level further downstream via
(1+1)-REMPI. Although no voltages are applied to the chip electrodes for these
measurements, the 50\micron\ high entrance and exit slits ensure that only
molecules within 50\micron\ from the surface of the chip can contribute to the
observed signal. The spectrum is shown and assigned in Figure~\ref{F:13CO}. The
width of the individual $M$-components of 400--500~kHz is attributed to
inhomogeneities of the magnetic field above the chip, induced by the metal
components of the chip-holder. It is important to realize that the rotational
transition can only be observed when the electric field of the chip is off; when
the electric fields are present above the chip while guiding or decelerating
molecules, the Stark broadening of each transition is many GHz, bringing the
vast majority of molecules out of resonance with the narrow band mm-wave
radiation. 

\subsection{Spectroscopic identification of CO molecules\label{sec:cleaning}}
In previous experiments in which we guided and decelerated metastable CO molecules on
a chip, we observed a background signal from CO molecules that were rather insensitive 
to the electric fields above the chip.\cite{Meek:2008p153003,Meek:2009p055024} We
tentatively attributed this background signal to metastable CO molecules in the
non-Stark-shifted $M$-levels of the upper $\Lambda$-doublet component of
$J=1$. With the mm-wave radiation we can now spectroscopically test whether this 
assignment is correct. For this, we irradiate the molecules on the chip with mm-waves 
that are resonant with a rotational transition from these $M$-levels to a low-field-seeking 
component of the $J=2$ level. Molecules in this low-field-seeking state will be deflected 
upwards by the exponentially decaying electric field above the chip, and will no longer pass 
through the 50\micron\ high exit slit at the end of the chip. Any molecule
undergoing this transition whose velocity does not match the trap velocity cannot be
captured by the traveling field minima or stably guided to the end of the chip. If our
previous assignment is correct, therefore, resonant mm-wave excitation will
result in a depletion of the background signal.

\begin{figure}
\centering
\includegraphics[width=0.47\textwidth]{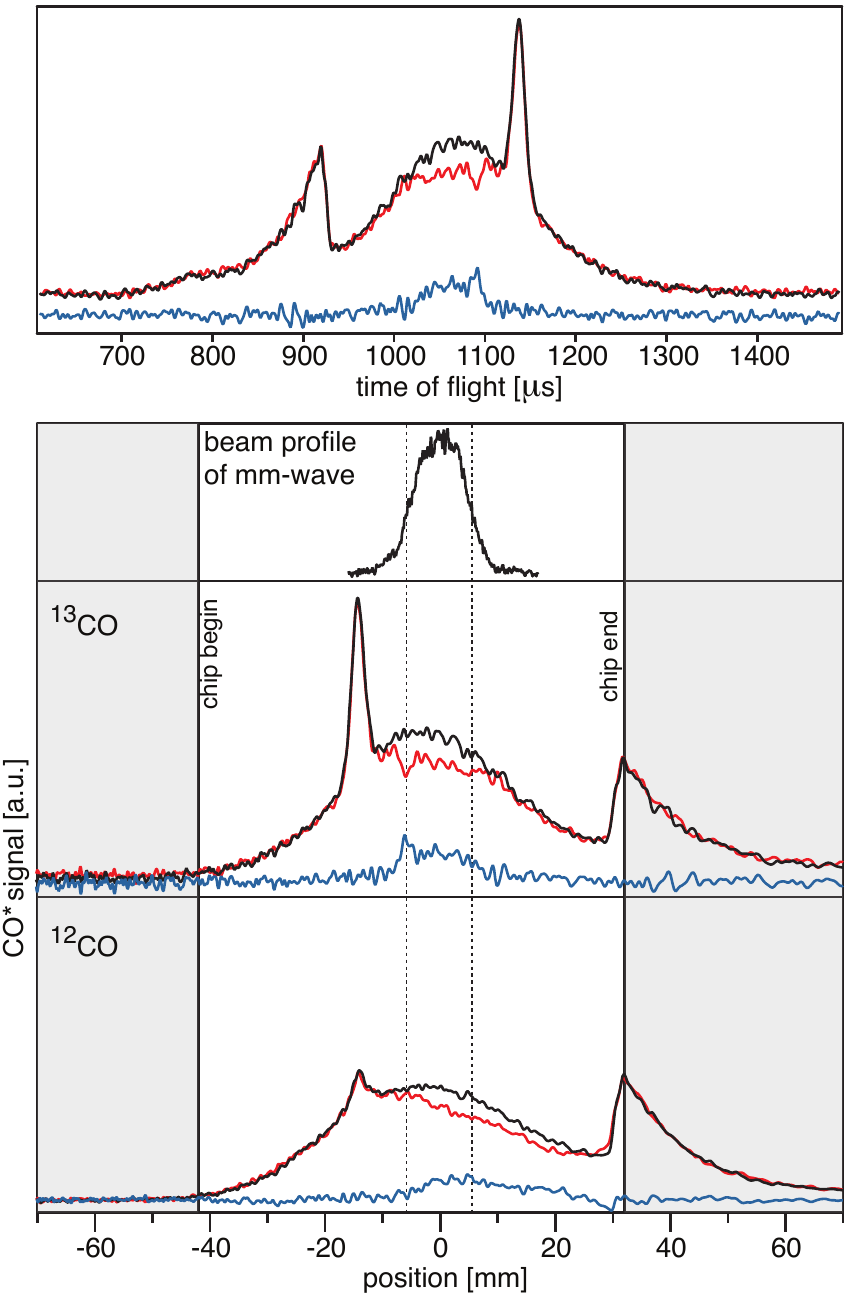}
\caption{Top: Measured arrival time distribution of metastable $^{13}$CO molecules 
guided over the chip at 300~m/s without (black curve) and with (red curve) mm-wave 
radiation applied, relative to the time of laser preparation. The difference between 
these two curves is plotted in blue. Main panel: Measured arrival time distribution,
shown as a function of the position of the CO molecules at the moment that the 
electric fields are applied to the chip. The position axis is relative to the center of the
spatial profile of the mm-wave beam, which is shown as well.\label{F:COClean}}
\end{figure}

For these experiments, the General valve is cooled to 140~K to reduce the mean
velocity of the CO beam to 300~m/s; the presently available electronics do not
allow guiding of the 
molecules on the chip at much higher speeds than this. The mm-source is resonant with the $J=2$, 
$F=3/2$, $M_F=3/2\leftarrow J=1$, $F=1/2$, $M_F=1/2$ transition in $^{13}$CO (indicated
with an asterisk in Figure~\ref{F:13CO}), promoting molecules from a non-Stark-shifted
level into a low-field-seeking one. Instead of state-selective detection of the metastable CO 
molecules via (1+1)-REMPI, the full velocity distribution of the beam of metastable CO 
molecules --- independent of the rotational state that they are in --- is monitored by time-resolved 
detection of the Auger electrons that are generated upon impact of these molecules on a gold 
surface.\cite{Meek:2009p055024} Measurements of the arrival time distribution of the $^{13}$CO
molecules are shown in the upper part of Figure~\ref{F:COClean}, both without (black curve) 
and with (red curve) the mm-wave radiation present; the difference between these 
measurements, showing the depletion in the background signal, is shown as well (blue curve).
The main peak in the signal, observed at 1140\microsec, results from CO molecules that are
stably guided over the chip in traps that move with a velocity of 300~m/s; from this the distance
from the laser excitation point to the Auger detector can be precisely determined as 342~mm. 
The abrupt cut-off in signal around 930\microsec\ reflects the abrupt switching on of the voltages
on the molecule chip; the signal at earlier times results from fast CO molecules that had already 
left the chip when voltages were applied and that are therefore unaffected. Most of the CO molecules 
in low-field-seeking states that are on the chip when the voltages are applied are deflected 
upwards and will thus no longer make it to through the 50\micron\ exit slit to the detector,
thereby explaining the abrupt signal decrease. It is clear from the measurement, however, that 
there is a prominent background signal remaining, and that this background signal can indeed be
partially depleted with the mm-waves. The depletion is expected to be at most half of the 
background signal since the CO molecules are distributed over two levels ($J=1$, $F=1/2$, 
$M_F=\pm1/2$) in the magnetic field, whereas only one of these is addressed by the mm-waves. 
The experimentally observed depletion is about 20\%, indicating that the population transfer to 
the $J=2$ level does not happen with unit efficiency. This could be explained by
a mismatch between the Rabi period and the interaction time or a mismatch between
the width of the mm-wave sectrum and the width of the absorption line. Moreover, some
rotationally excited molecules might not be deflected strongly enough and might
still make it through the exit slit. The latter will certainly hold for those
rotationally excited molecules that are already close to the exit of the chip
when the electric fields are switched on.

To aid in the interpretation of the measured arrival time distributions, the time
scale can be converted to a position scale, showing where the CO molecules 
that contribute to a particular signal were located at the time that the voltages 
on the chip were switched on. These spatial distributions are shown in the lower 
part of Figure~\ref{F:COClean}, together with the profile of the mm-wave beam 
on the chip; the zero of the position scale is arbitrarily taken at the center of the 
mm-wave beam. Only molecules that are in, or have already passed, the mm-wave 
beam before the voltages on the chip are switched on can be rotationally excited;
when the electric fields on the chip are on, the rotational transitions are Stark-broadened
too much. A clear depletion of the background signal is indeed only observed for 
those molecules that interacted with the mm-waves prior to switching on the
electric fields.

In the lower part of Figure~\ref{F:COClean} similar measurements are shown
for $^{12}$CO, spectroscopically identifying the metastable CO molecules that
contribute to the background signal as molecules in $M=0$ levels. The large
difference in the efficiency of guiding either $^{13}$CO or $^{12}$CO on the chip, 
is evident from these measurements. 

\subsection{Population transfer between trapped levels}

In the tubular electric field traps above the chip, low-field-seeking CO molecules 
are exposed to electric field strengths that range from zero up to several kV/cm. 
As the molecules sample a large fraction of these field strengths on a\microsec\ 
time-scale and as the Stark shifts in the most strongly confined $M$-components of the 
$J=1$ and $J=2$ level are sufficiently different, the traps must be temporarily 
turned off to induce a rotational transition between these levels. After this, the traps 
can be turned on again and the CO molecules can be recaptured. For the overall 
efficiency of this scheme, the motion of the molecules relative to the center of the 
moving trap is important. In our experiment, the molecules have a velocity spread 
inside the trap of a few m/s, limiting the maximum time during which the trap
can be switched off to several\microsec. From the experiments in the free beam it is
seen, however, that near-complete population transfer between the rotational 
levels in CO can be achieved by Rabi flopping in less than 3\microsec; guiding, releasing, 
transferring population and re-catching the molecules thus seems possible.

\begin{figure}
\centering
\includegraphics[width=0.47\textwidth]{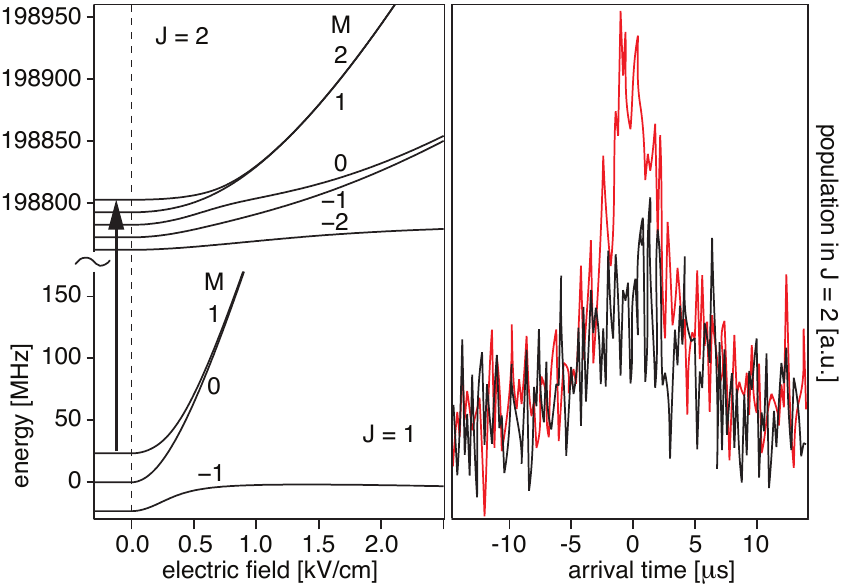}
\caption{Left: Energy level scheme of $^{12}$CO as a function of electric field in
the presence of a magnetic field of 50~Gauss perpendicular to the electric field.
Right: Arrival time distribution of metastable CO molecules in the $J=2$ level, as
measured via (1+1)-REMPI detection. The CO molecules have been prepared
in the $J=1$ level, and the signal is shown without (black trace) and with (red trace)
resonant mm-wave radiation present. The time zero is when molecules with a 
velocity of 300~m/s should arrive.\label{F:flip}}
\end{figure}

The arrival time distributions of metastable CO molecules subjected to such a 
procedure are shown in Figure~\ref{F:flip}. The $^{12}$CO molecules are prepared
in the $J=1$ level and guided to the center of the chip in traps that move at a speed 
of 300~m/s. When the molecules arrive near the center of the chip, the electric fields 
are switched off and the mm-wave source drives the $J=2$, $M=2\leftarrow J=1$, 
$M=1$ transition, as schematically indicated in the panel on the left. During this time, 
the molecules continue to move in the $z$-direction with a mean speed of 300~m/s, 
but they are no longer confined, and some of them are therefore not recaptured when 
the trapping voltage is turned on again 4\microsec\ later. Those molecules 
that are re-captured are guided to the end of the chip and fly into the ionization detection
region. The red curve in Figure~\ref{F:flip} shows the yield of molecules in the $J=2$ level, 
after interaction with the mm-wave field, as a function of time; the zero on the time axis is 
when molecules with a velocity of 300~m/s should arrive. The black curve shows the 
result when the mm-wave radiation is off; this background signal results from $J=1$ 
molecules that are non-resonantly ionized by the laser that is tuned to probe the $J=2$ 
level via the $N'=0$ level in the $b$-state.

\section{Conclusion}
The combination of (i) laser preparation of molecules in a single rotational level 
at a well-defined time and position, (ii) tunable narrow-band mm-wave 
radiation that can transfer the population to another rotational level at a known distance 
downstream, and (iii) state-selective detection of the molecules at a known delay 
and at a known position, yields unique possibilities. It has been used here for
a detailed {\it in situ} characterization of the mm-wave beam, which is non-trivial
otherwise at these wavelengths. Moreover, we have used it to experimentally 
record the time-dependence of the population transfer in a true two-level system,
not only in the case of Rabi-cycling but also under conditions where rapid adiabatic 
passage occurs; the measurements shown in Figure~\ref{F:rabi} are a text-book
example for these processes. 
The freely propagating mm-wave beam produced by the Armadillo can be
conveniently coupled to the molecules on the chip. We have used 
this to spectroscopically verify that the background signal observed in earlier experiments 
on guiding CO molecules on a chip originates from molecules in $M=0$ levels, as
anticipated. Last but not least, CO molecules in the $J=1$ level have been guided 
on the chip and have then been rotationally excited with the mm-wave radiation to 
the $J=2$ level, in which they have been recaptured on the chip.

Characterizing the electric field of the mm-wave beam above the chip is more
challenging than for the free beam, in particular because of the magnetic field
inhomogeneity at the surface of the chip. More generally, whenever a high 
spectroscopic resolution is required in the experiment, this inhomogeneity has 
to be minimized. The reflection of the mm-waves from the holder of the chip as
well as from the electrodes has to be considered in the design phase of the next
generation of microchips.

\section{Acknowledgment}

This work has been funded by the European Community's Seventh Framework Program
FP7/2007-2013 under grant agreement 216 774, and ERC-2009-AdG under grant agreement
247142-MolChip. We acknowledge fruitful discussion with Boris
Sartakov. G.S. gratefully acknowledges the support of the Alexander von Humboldt-Stiftung.

\bibliographystyle{gams-notit-nonumb}
\bibliography{biblio}

\end{document}